\documentclass[twocolumn,showpacs,showkeys,preprintnumbers,amsmath,amssymb]{revtex4}

\usepackage{epsfig}

\begin{document}


\title{An intensity-expansion method to treat non-stationary time series: \\ an application to the distance between prime numbers.}

\author{N. Scafetta,$^{1,2}$ T. Imholt,$^3$  J.A. Roberts,$^3$ and B.J. West.$^{1,2,4}$}
\affiliation{$^{1}$ Physics Department, Duke University, Durham, NC 27708. }
\affiliation{$^{2}$ Pratt School of EE Department, Duke University,  P.O. Box 90291, Durham, NC 27708. }
\affiliation{$^{3}$ Center for Nonlinear Science, University of North 
Texas,   P.O. Box 311427, Denton, Texas 76203-1427 }
\affiliation{$^{4}$ Mathematics Division, Army Research Office, Research Triangle Park, NC 27709. }

\date{\today}

\begin{abstract}
We study the fractal properties of the distances between consecutive primes. The distance sequence is found to be well described by a non-stationary exponential probability distribution. We propose an intensity-expansion method to treat this non-stationarity and we find that the statistics underlying the distance between  consecutive primes  is Gaussian and that, by transforming the distance  sequence into a  stationary one,  the range of Gaussian randomness of the sequence increases.   
\end{abstract}

                           

\maketitle

\section{Introduction }
Many complex systems have been  empirically shown to be characterized by scale invariance  \cite{2Mandelbrot} and the correct evaluation of the scaling 
exponents is of fundamental importance to assess if universality 
classes exist \cite{stanley}. However, a correct interpretation of the results of the statistical analysis often requires an examination of the stationarity of the time series under study. Stationarity  means that the probability distribution for  a given process is invariant under a shift of the time origin \cite{priestley}. In particular,  if $\{X_i\}$ is  a time series collecting the measure of an physical observable, the probability density function (pdf) $p(X,t)$ is stationary if it remains invariant by shifting the origin of time. 

However, time series are often not stationary. Their pdf changes in time and  caution should be taken to analyze non-stationary time series \cite{scaff,stan1,stan2}. Here, we suggest a method to reduce the non-stationarity of a time series through an intensity-expansion method that transforms the original non-stationary time series into a new one that  satisfies the stationary condition.   The idea is to study the evolution of the pdf of the time series $\{X_i\}$ and to  look for a law that describes the non-stationarity of the  pdf $p(X,t)$ of the time series.  Finally, we transform the original data into a new time series $\{Y_i\}$ such that Y=F(X,t). The new time series $\{Y_i\}$ is described by a pdf $p(Y)$ such that 
\begin{equation}\label{pptras}
p(Y) dY= p(X,t)~dX~.
\end{equation}   
Eq. (\ref{pptras}) assures that the transformed time series $\{Y_i\}$ no longer depends on  time  and, therefore,  is now stationary. So, the latter time series  may be studied via  conventional stochastic methods. 

As an example, we apply our method to a well known non-stationary time series in mathematics, the distance ( or waiting time) between two consecutive prime numbers. Fig. 1 shows the first 1000 data points of a 110 million point time series. The main source of non-stationarity is due to the fact that the average distance between two consecutive primes tends to increases as the numbers increase. In fact, Gauss, in an  1849 letter to the astronomer Hencke, stated that he had found a function  that approximately gives the number of primes  up to the number N, that is, the well-known log-integral function \cite{riemann}
\begin{equation}\label{logint}
Li(N)=\int\limits_{0}^{N} \frac{dt}{\log(t)}~.
\end{equation}
Eq. (\ref{logint}) shows that the distance between consecutive primes is non-stationary because its average $\Delta(N_0,N)$ in the range $ [N_0,N_0+N]$ depends on the origin $N_0$ according to the approximate relation
 \begin{equation}\label{apprel}
\Delta(N_0,N)\approx \frac{N}{Li(N_0+N)-Li(N_0)}~.
\end{equation} 
The non-stationary is due to the fact that the log-integral function, $Li(N)$, is not linear in $N$.

This problem is represented in many different ways and the most compact form of it can be found in the Riemann hypothesis \cite{riemann}.  This problem, in the opinion of many mathematicians, is probably  today's most important unsolved problem in pure mathematics.  Many references to the Riemann hypothesis and its early history  can be found in Landau \cite{landau} and Edwards \cite{edwards}.

However, Eq. (\ref{logint}) gives only approximate information about the mean value and does not say anything about the properties of the fluctuations of the distance between consecutive primes. Therefore, several questions arise, including the  particular form of  the pdf of the distance between primes;   how the pdf evolves in the time and wheather the fluctuations  contain some form of long-range correlation.  Finally, we investigate  the role played by the non-stationary properties of the time series   and how to handle these properties. 

In Sec. II we review two complementary statistical methods used to detect scaling properties in a time series. In Sec. III we analyze the sequence of distance between primes  and in Sec. IV we draw some conclusion.

\section{Scaling analysis}

In the analysis of the distance between prime numbers we  make use of pdf analysis and of two complementary scaling analysis methods: the  standard deviation analysis (SDA) and the diffusion entropy analysis (DEA) \cite{thermodynamicofsocialprocesses,dea4,dea6,scalingdetection,dfa}. The need for using these two methods to analyze the scaling properties of a time series is to discriminate  the stochastic nature of the data: Gaussian or  L\'evy \cite{dea4}.  

We analyze the scaling exponents of the diffusion process generated by a time series.  
The SDA allows one to determine the scaling exponent of the standard deviation  of the diffusion pdf $p(x,t)$ with time, that is usually called the Hurst exponent $H$. Whereas DEA allows one to determine the scaling exponent  $\delta$ of the same pdf of the   diffusion process under study. Finally we compare $H$ and $\delta$. In the particular case in which the data are characterized by  Gaussian statistics, it is possible to prove that $H=\delta$ . If $H\neq \delta$ the scaling presents anomalous behavior. Random  or uncorrelated Gaussian noise is characterized by $H=\delta=0.5$. If the noise is long-range correlated we have $0<H<0.5$ for antipersistent noise and $0.5<H<1$ for persistent noise \cite{2Mandelbrot}. We stress that the fractal correlation properties detected by these techniques are accurate only if the time series used to generate the diffusion process is stationary. In fact,  non-stationarity may be mistaken for  correlations because they  yield  a  memory that may generate an anomalous behavior in the scaling exponents \cite{thermodynamicofsocialprocesses}. So, some caution should be taken in treating  non-stationary signals.

According to the prescription of Ref. \cite{dfa}, we interpret the 
numbers of 
a time series as generating diffusion fluctuations and we shift our 
attention 
from the time series to the pdf 
$p(x,t)$, 
where $x$ denotes the  variable collecting the fluctuations. The 
scaling 
property of the pdf takes
\begin{equation}\label{scafun12}
p(x,t) = \frac{1}{t^{\delta}}~F\left( \frac{x}{t^{\delta}}\right)~,
\end{equation}
where $\delta$ is the scaling exponent. Practically,
let us consider a sequence of $N$ numbers
\begin{equation}
    \xi_{i} ,    \quad i = 1,  \ldots ,N.
    \label{thesequenceundestudy}
    \end{equation}
    The goal is to establish the possible 
    existence of  scaling, either normal or anomalous in this sequence. 
First of all, let us select  an integer number
    $t$, fitting the condition $1 \leq t < N$. 
This integer number will be referred  us as the ``diffusion time''. 
For any 
given 
    time $t$ we can find $M(t)=N - t +1$ sub-sequences defined by
    \begin{equation}
        \xi_{i}^{(s)} \equiv \xi_{i + s}, \quad with  \quad s = 0,  
\ldots ,  
N-t.
        \label{multiplicationofsequence}
        \end{equation}
        For any of these sub-sequences we build up  a diffusion 
trajectory, 
$s$, defined by the position        
        \begin{equation}
    x^{(s)}(t) = \sum_{i = 1}^{t} \xi_{i}^{(s)} 
    = \sum_{i = 1}^{t} \xi_{i+s}.   
        \label{positions}
        \end{equation}

The direct evaluation of the variance is probably the most natural method 
of 
variance scaling detection. All
trajectories start from the origin $x(t=0)=0$.  With increasing time 
$t$, the 
sub-sequences generate a diffusion
process. At each time $t$, it is possible to calculate  the
standard deviation of the position of the $M(t)$  sub-sequences with 
the 
well known expression for the standard deviation:
\begin{equation}\label{varvar33}
D(t) =\sqrt{\frac{ \sum _{s=0}^{N-l}\left[x^{(s)}(t)-\overline{x}(t) 
\right]^2}{M(t)-1}},
\end{equation}
where $\overline{x}(t)$ is the average of the positions of the
$M(t)$ sub-trajectories at time $t$. The exponent $H$ is defined for a scaling diffusion process by 
\begin{equation}\label{scaliH}
D(t)\propto t^H ~.
\end{equation}

The DEA, is based upon the following algorithm. We partition 
the $x$-axis into cells of size 
       $\epsilon(t)$ and label the 
       cells by $i=1,2,\dots$. We count how many particles are found in the same cell at 
a given time $t$. We denote this number by $N_{i}(t)$. Then 
       we use this number to determine the probability that a particle 
       can be found in the $i$-th cell at time $t$, $p_{i}(t)$, by 
means 
       of
       \begin{equation}
        p_{i}(t) \equiv  \frac{N_{i}(t)}{M(t)} .
        \label{probability}
        \end{equation}
        At this stage the entropy of the diffusion process at  time 
$t$
        can be determined and reads
\begin{equation}
 S(t) = - \sum_{i} p_{i}(t) ~\ln [p_{i}(t)]~.
\label{entropy}
        \end{equation}
        The easiest way to proceed with the 
choice of the cell size, $\epsilon(t)$, is to assume it to be a 
fraction of 
the square root of 
the variance of the 
fluctuations $\xi(i)$, and consequently independent of $t$. 
If the scaling condition  (\ref{scafun12}) holds true, it is easy 
to 
prove that 
\begin{equation}\label{scafun14}
S(t) = A + \delta~  \ln (t) ~,
\end{equation}
        where, 
in the continuous approximation,
        \begin{equation}
        A \equiv -\int_{-\infty}^{\infty} dy \, F(y) \, \ln [F(y)]~,        
\label{ainthecontinuouscase}
        \end{equation}
with  $y = x/t^{\delta}$.
The scaling in Eqs. (\ref{scaliH}) and  (\ref{scafun14}) determine the 
exponents 
$H$ and $\delta$, respectively.

The comparison between DEA and some other traditional techniques of scaling detection, among them the detrended fluctuation analysis \cite{dfa} and a wavelet variance based method, for both L\'evy and Gauss statistics, are made in  Ref. \cite{scalingdetection}.

\section{Prime number analysis}

The time series here under study is the  distance $X_i$ between two consecutive prime numbers. For example, for the first 6 prime numbers ($p_i=2,3,5,7,11,13\dots$) the distances are $X_i=p_{i+1}-p_{i}=1,2,2,4,2\dots$ with the index $i=1,2,\dots$.    Fig. 1 shows the distance $X_i$ of the first 1000 prime numbers. The entire time series that we analyze contains almost 110 million distances that correspond to all prime numbers between 1 and $2^{32}$, that is, the largest integer handled by a 32 bits computer.  

First, we analyze the stationarity of our time series. To do this we partition the entire time series into 110 consecutive subsets of 1 million data points each and study the distribution of the data for each subset.
Fig. 2 shows the distribution of the  distance $X_i$ of the first and last subset. The figure shows that the distribution of the data are well fitted by an exponential distribution of the kind
\begin{equation}\label{expdis}
p(X,n)= A(n) \exp{\left[-k(n) X \right]}
\end{equation}
where the integer $n$ is the index of the subset and  $\tau(n)=1/k(n)$ is the characteristic distance between two primes.  The dependence of the characteristic time $\tau(n)$ on the subset index $n$ indicates that the time series of distance $X_i$ between two consecutive prime numbers is not stationary and some caution should be taken in the analysis of this time series. We observe that Kumar {\it et al.} recently noticed that this distribution follows an exponential form \cite{kumar}, but they did not notice that this esponential form is not stationary.  

The non-stationarity of the  distances between primes is depicted in Fig. 3. We plot the value of the exponential fitting parameters $k(n)$ for all 110 subsets. We see that  the values $k(n)$ decrease and fluctuate around a logarithmic curve of the type
\begin{equation}\label{logcurv}
k(n)=a+b\log(n)+O(\log(n)^2)
\end{equation}
where the best least-square fitting parameters are $a=0.074\pm0.002$ and $b=-0.0041\pm0.0002$. The decrease of $k(n)$ implies that the average distance between two primes, that is related to $\tau(n)$, tends to grow with the size of the numbers themselves. Fig. 4 shows the values of the fitting coefficients $A(n)$ of Eq. (\ref{expdis}) and Fig. 5 shows the mean value of the distance between consecutive primes for each one million data subset compared with the approximated mean value given by 
 \begin{equation}\label{appr7el}
\Delta(n)\approx \int\limits_{0}^{\infty }X~p(X,n)~dX=\frac{A(n)}{k(n)^2}~,
\end{equation} 
that is obtained by using the pdf (\ref{expdis}).

Eq. (\ref{logcurv}) expresses only the linear approximation to the non-stationary behavior of the  distances between primes because  $k(n)$ cannot become negative since the characteristic time, $\tau(n)$, must be positive. Therefore$k(n)$ can, at most,  only asymptotically converge to zero. However, Fig. 3 gives some indication of the goodness of the linear fitting of Eq. (\ref{logcurv}) in the range of numbers that we consider here.

Eqs. (\ref{expdis},\ref{logcurv}) also suggest how to reduce the non-stationarity of the distance series between primes. In fact, we can use the time series of $X_i$ and the equation (\ref{logcurv}) to define an intensity-expanded series $Y_i$, with $i=1,2,\dots$ in the following way:
\begin{equation}\label{nonstatimeser}
Y_i=k(10^{-6}~i)~X_i ~.
\end{equation} 
Note  that we substitute the integer $n$ with $10^{-6}~i$ because in our notation  $n$ represents the index of the $n-th$ subset of 1 million distances between primes. In fact, it is easy to realize that according to  the Eqs. (\ref{expdis}) and (\ref{logcurv}) the new time series  $Y_i$ is more stationary than the previous series of  distances $X_i$ because 
the values $k(n)$ for the time-expanded time series will fluctuate around the constant value  $k(n)\simeq 1$.  So,  each subset of 1 million data of the time-expanded series $Y_i$  are distribute around  the same probability density function $p(Y)\approx \exp\left[-Y \right]$.  This property assures an approximate stationarity of the intensity-expanded series $Y_i$ and allows the adoption of  analyses that require the stationarity of the dataset. So, we conclude that in the particular case of the distance between primes the transformation (\ref{nonstatimeser}) satisfies the condition expressed by Eq. (\ref{pptras}).

Figs. 6 and 7 show respectively the DEA and SDA applied to both datasets of the original distances $X_i$ and  of the time-expanded series $Y_i$.  Both figures show that by increasing the diffusion time $t$ both scaling exponents $\delta$ and  $H$ change from the value $\delta=H=0.5$ at short range to $\delta=H=0.88$ at long range. The fact that the two scaling exponents are the same indicates that the data are consistent with Gaussian statistics. 

Figs. 6 and 7 allow us to conclude that  the statistics of  the distance between primes can be considered well described by random noise plus a temporal amplification due to the non-stationary mechanism indicated in Figs. 2-5 and described by Eqs. (\ref{logcurv}) and (\ref{appr7el}). In fact, both DEA and SDA of the distances between primes (shown in the figure by the curves with triangles) have the scaling exponents $\delta=H=0.5$  until $t\simeq 50$ and have $\delta=H=0.88$ for $t>50$. With these results it is possible to mistake the distances between primes for a random signal that is practically uncorrelated at short range and strongly correlated at long range.  However, the curves concerning the DEA and SDA applied to the intensity-expanded series $Y_i$ given by (\ref{nonstatimeser}) suggest a different interpretation. In fact, with a simple time transformation finalized to reduce the non-stationary properties of the original time series,  it is possible to extend the uncorrelated range from a few decades, that is, $t<50$ to $t<10000$ as  Fig. 6 and 7 clearly show (curves with circles).  Thus, the results shown in Fig. 6 and 7 suggest that the high value of the scaling exponents, $\delta=H=0.88$, at high range, at least between $50<t<10000$, is not a manifestation of long range fractal noise but of the non-stationarity of the original time series.  Perhaps, with better fitting procedures and an higher order fitting function (\ref{logcurv}), it will be possible to extend the range of uncorrelated noise further.

\section{Conclusion}
In this paper we have introduced a  method to transform a  non-stationary time series into a new time series that better satisfies  the stationary condition. We study the evolution of the probabilistic structure of the original time series $\{X_i\}$, that is,  we study the non-stationary pdf $p(X,t)$ of the original time series and, finally, we transform the original data into a new time series $\{Y_i\}$ through an intensity-expansion mechanism such that Y=F(X,t) and such that the pdf $P(Y)$ is independent of $t$.

The DEA and SDA  techniques are used to study the statistics of the distances between prime numbers. From this study it is seen that the distance between primes generates a non-stationary time series due to an increasing mean distance between primes. This non-stationarity  is responsible for the apparently  persistent fractal nature of the time series at large time-range. However,
upon removal of the nonstationary components of this time series, the  near neighbor primes show a great deal of randomness for a much longer range.  This may imply that a deeper understanding of the distribution of primes along the real axis may be gained by conducting a similar study with much larger datasets.  This technique, does however suggests  that the distribution of primes in consistent with  Gaussian statistics.

--------\newline
{\ {\large \textbf{Acknowledgment:}}}\newline
N.S. thanks the Army Research Office for support under grant DAAG5598D0002.

\newpage

\begin{figure*}
\epsfig{file=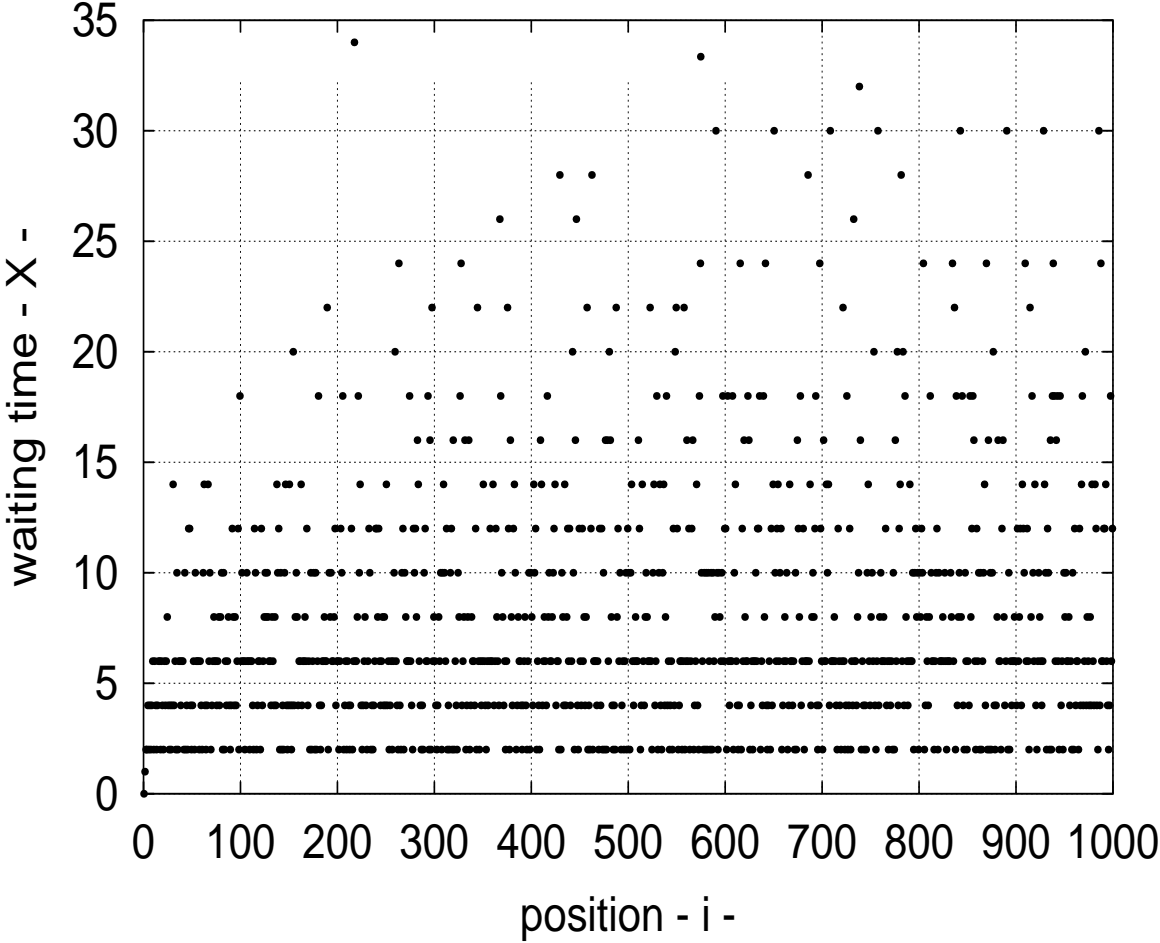,height=16cm,width=13cm,angle=-90}
\caption{The distance $X_i$ between of the first 1000 consecutive prime numbers.}
\end{figure*}

\begin{figure*}
\epsfig{file=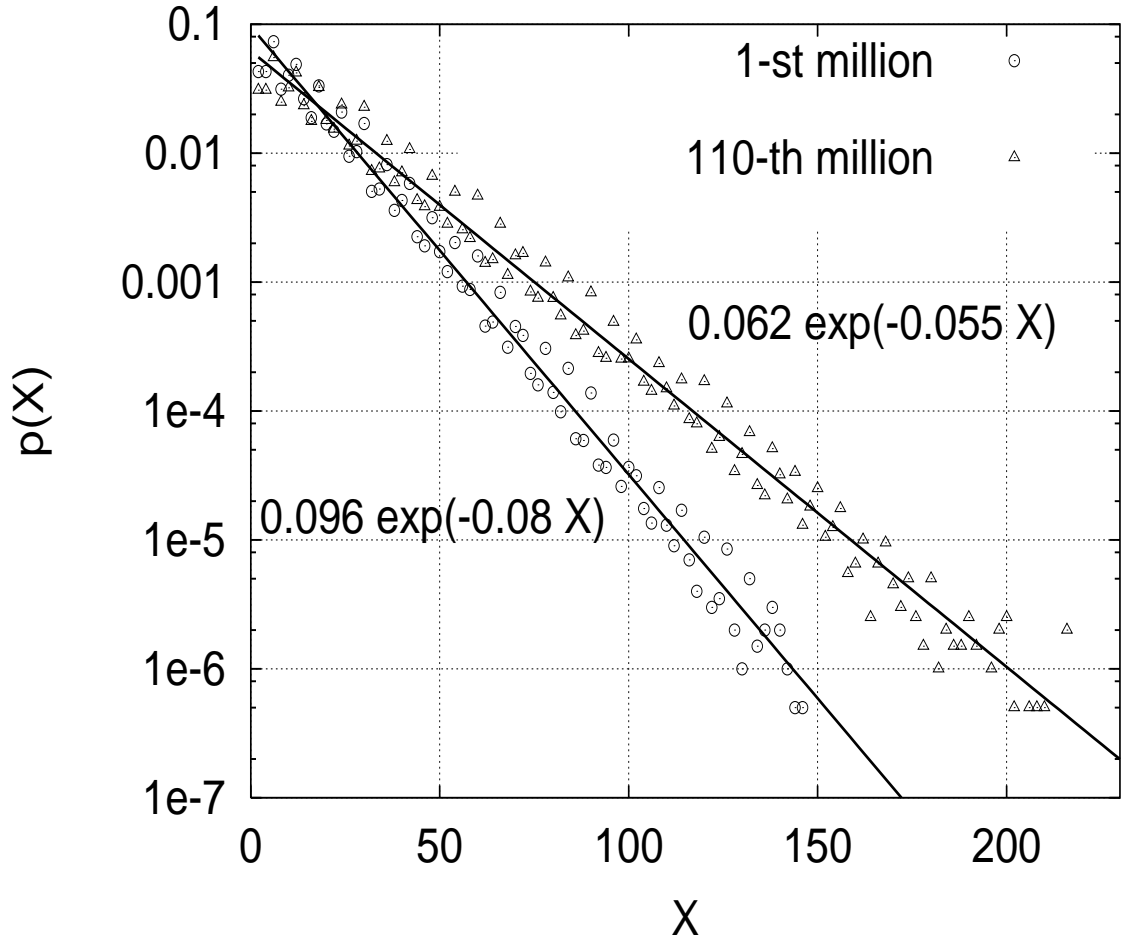,height=16cm,width=13cm,angle=-90}
\caption{Distribution of the  distance $X_i$ of the first subset of 1 million data and for the 110$-th$  subset. The distribution are fitted by an exponential distribution (\ref{expdis}).}
\end{figure*}

\begin{figure*}
\epsfig{file=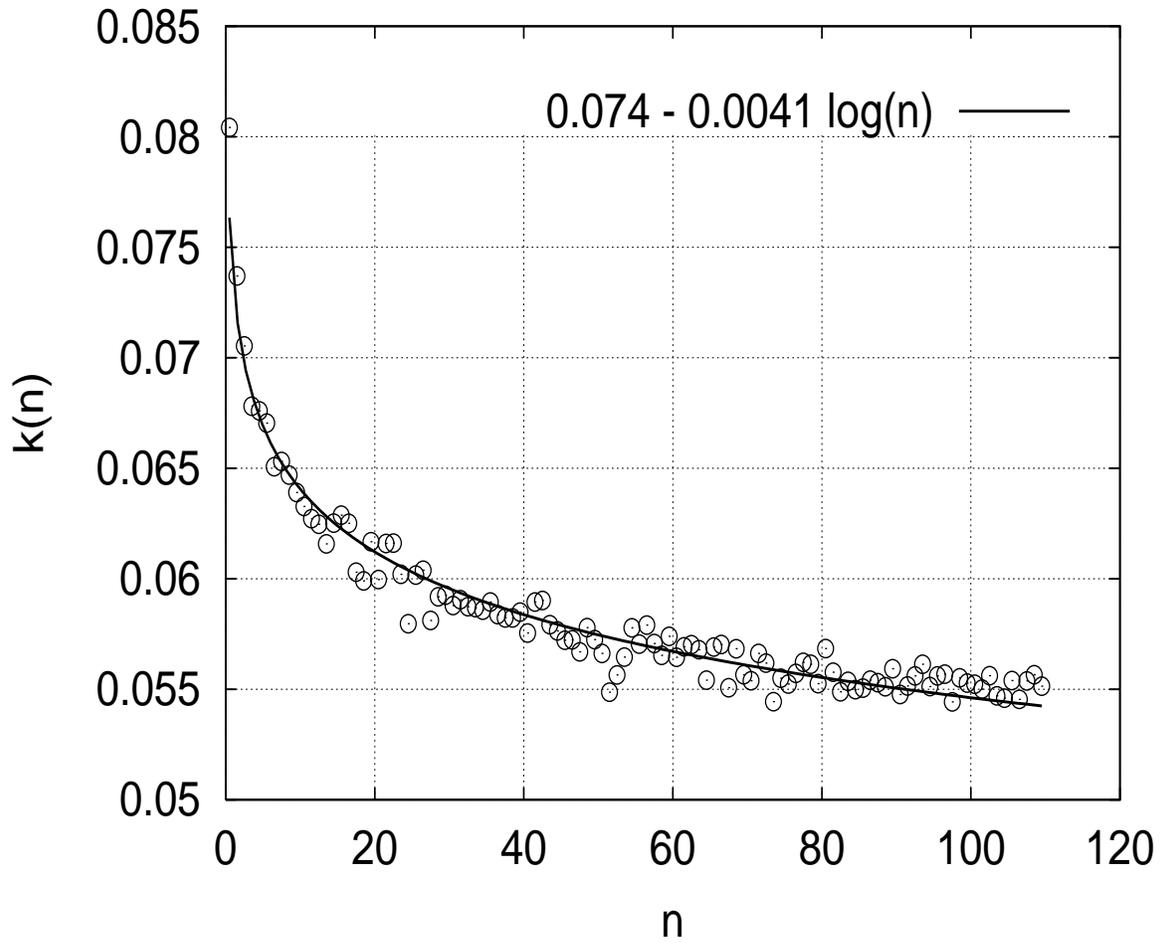,height=16cm,width=13cm,angle=-90}
\caption{The exponential fitting parameters $k(n)$ for all 110 subsets the  distance $X_i$.}
\end{figure*}

\begin{figure*}
\epsfig{file=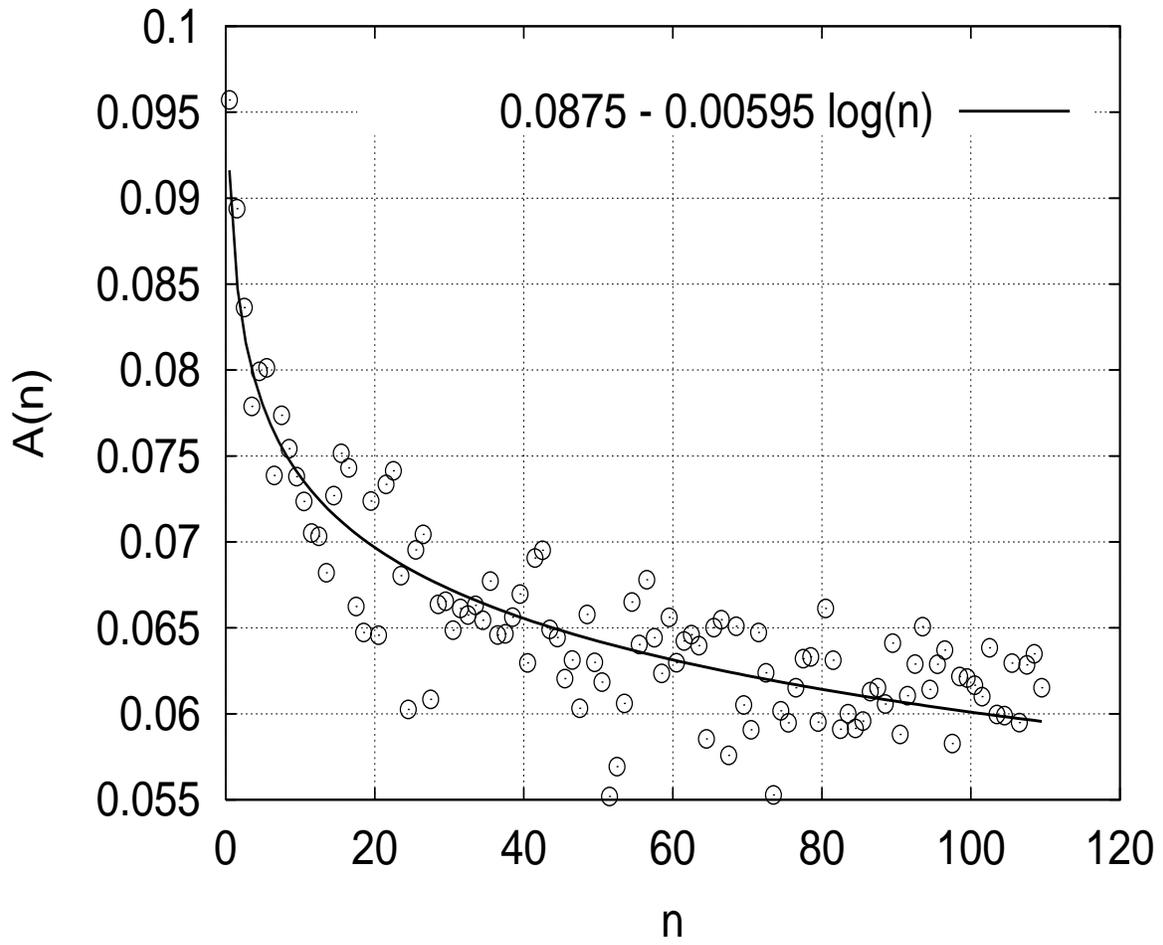,height=16cm,width=13cm,angle=-90}
\caption{Fitting coefficient $A(n)$ of pdf (\ref{expdis}).}
\end{figure*}

\begin{figure*}
\epsfig{file=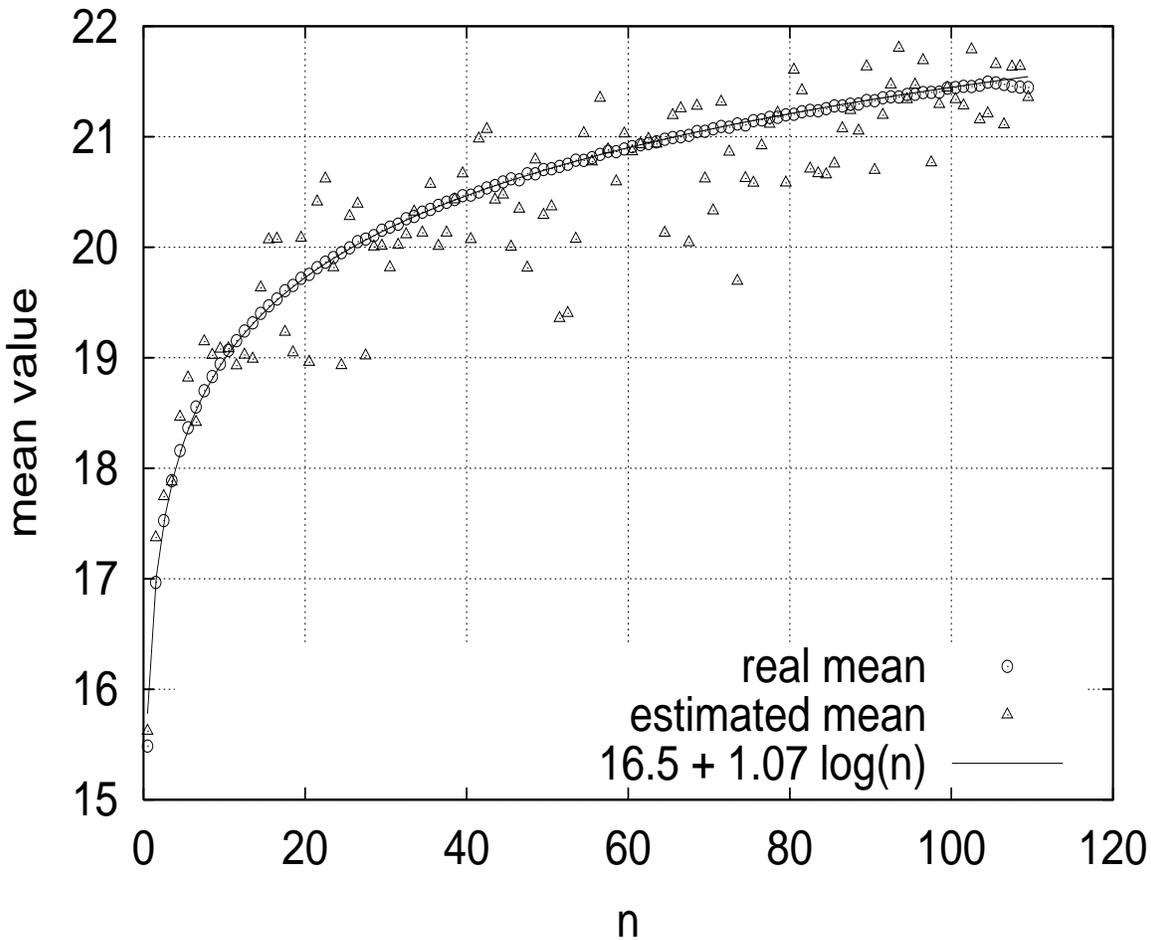,height=16cm,width=13cm,angle=-90}
\caption{The mean value of the distance between consecutive primes for each one million data subset (circles) compared with the approximated mean value given by  Eq. (\ref{appr7el}) (triangles). The error is almost 10\%.
 }
\end{figure*}

\begin{figure*}
\epsfig{file=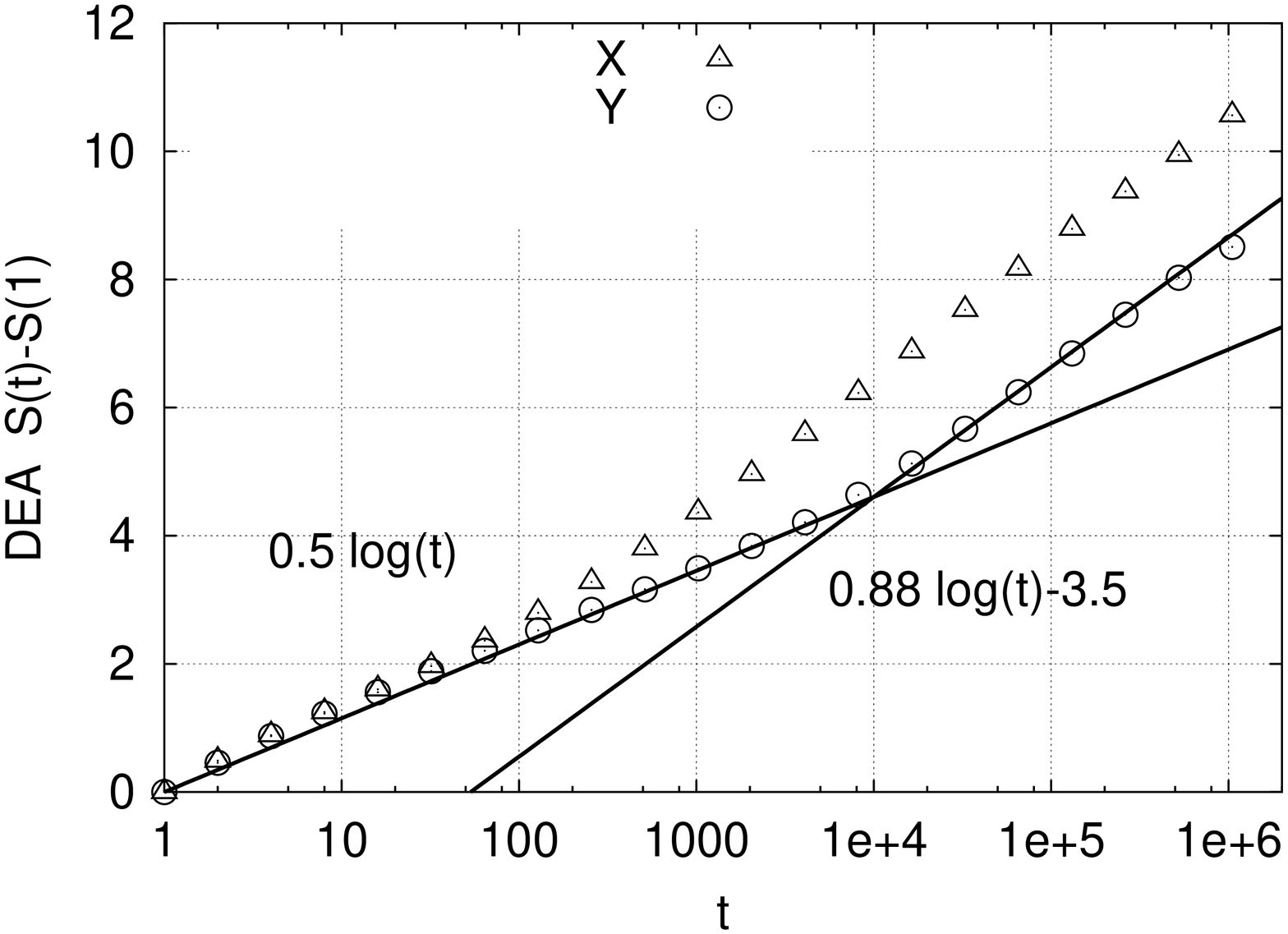,height=16cm,width=13cm,angle=-90}
\caption{DEA of the original distance dataset $X_i$ (triangles) and of the time-expanded dataset $Y_i$ (circles).}
\end{figure*}

\begin{figure*}
\epsfig{file=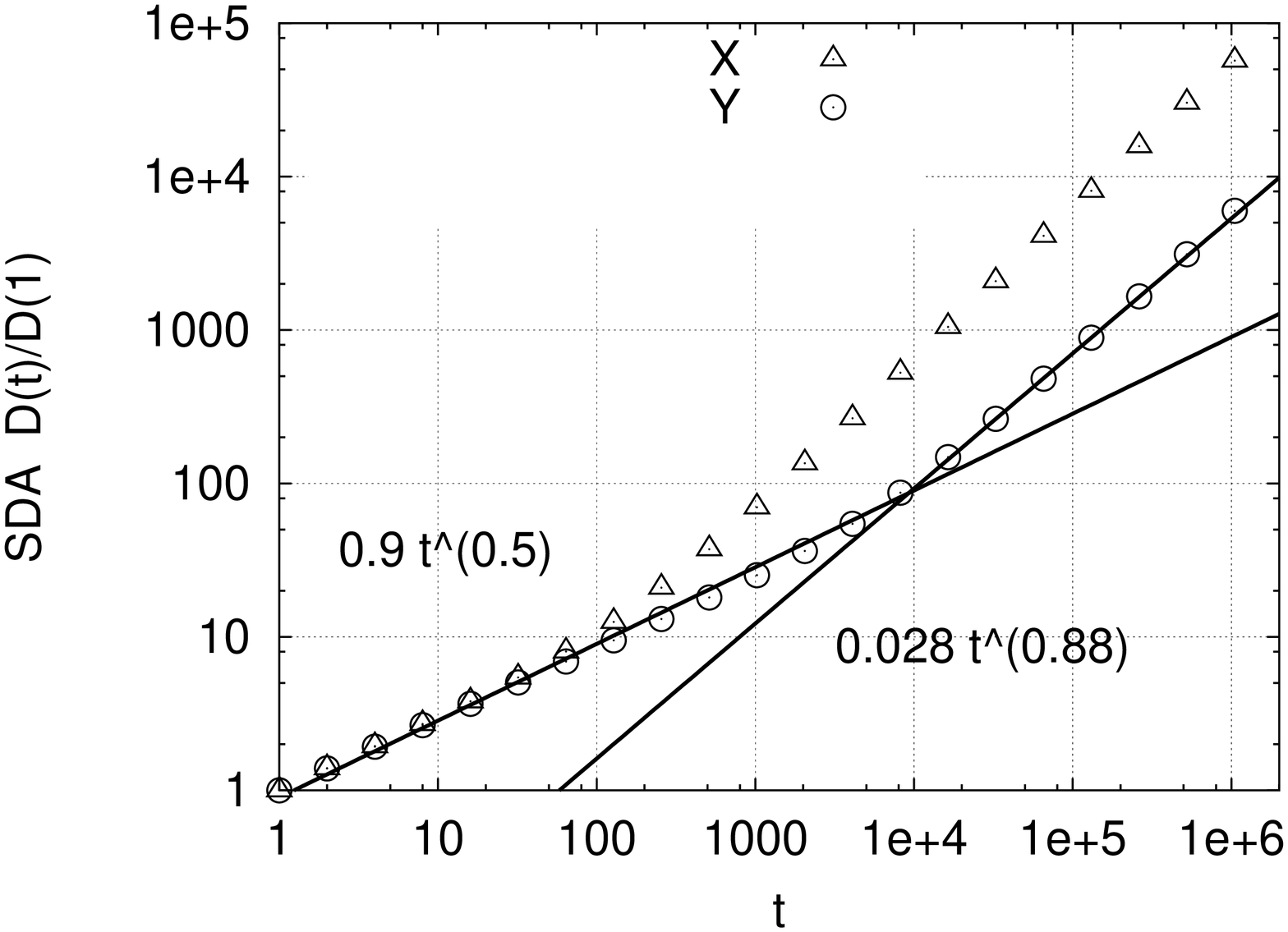,height=16cm,width=13cm,angle=-90}
\caption{DSDA of the original distance dataset $X_i$ (triangles) and of the time-expanded dataset $Y_i$ (circles).}
\end{figure*}

\end{document}